\documentclass[prd,twocolumn,tightenlines,superscriptaddress,nofootinbib,
showpacs]{revtex4}
\usepackage{amssymb,latexsym}
\usepackage{amsmath,amsbsy,bbm}
\usepackage{epsfig,bm}
\usepackage{graphicx,comment}
\usepackage{color}
\usepackage{soul}
\usepackage[normalem]{ulem}
\begin{document}

\title{Universal scaling of three-dimensional dimerized quantum antiferromagnets on bipartite lattices}

\author{D.-R. Tan}
\affiliation{Department of Physics, National Taiwan Normal University,
88, Sec.4, Ting-Chou Rd., Taipei 116, Taiwan}
\author{F.-J. Jiang}
\email[]{fjjiang@ntnu.edu.tw}
\affiliation{Department of Physics, National Taiwan Normal University,
88, Sec.4, Ting-Chou Rd., Taipei 116, Taiwan}

\begin{abstract}
Using the first principles quantum Monte Carlo (QMC) calculations,
we investigate the previously established universal scaling between
the N\'eel temperature $T_N$ and the staggered magnetization density
$M_s$ of three-dimensional (3D) dimerized quantum antiferromagnets.
Particularly, the calculations are done
on both the stacked honeycomb and the cubic lattices. In addition to 
simulating models with two types of antiferromagnetic couplings (bonds)
like those examined in earlier studies, here a tunable 
parameter controlling the strength of third type of 
bond is introduced. Interestingly,
while the data of models with two types of bonds obtained here fall on top
of the universal scaling curves determined previously, the effects due to  
microscopic details do appear. Moreover, the most striking result suggested
in our study is that with the presence of three kinds of bonds 
in the investigated models, the considered scaling relations between $T_N$ and
$M_s$ can be classified by the coordinate number of the underlying lattice
geometries. The findings presented here broaden the applicability of the 
associated classification schemes formerly discovered. In particular, 
these results are not only interesting from a theoretical point of view, 
but also can serve as useful guidelines for the relevant experiments.   
         
\end{abstract}
\vskip-0.25cm

\maketitle

\section{Introduction}
\vskip-0.1cm

Finding relations among quantities which  
are universal, namely being valid for various systems is a fascinating task
in the physical world. Moreover, to be able
to classify these relations are crucial and important
considering its great potential applications in
the relevant experiments. The critical exponents
of second order phase transitions is one of
such examples \cite{Nig92,Car10,Sac11}. For instance, for three-dimensional (3D)
classical Heisenberg model and any two-dimensional (2D) dimerized 
quantum spin systems, when the relevant phase transitions occur 
in these systems, their associated critical exponents all have the same 
numerical values \cite{Cam02,Pel02}. Because these models have either 
$O(3)$ or $SU(2)$ symmetry, this universality class
is called the $O(3)$ universality class in the literature. 
Other models having different symmetries and dimensions, such as 3D Ising model 
or 2D classical XY model, belong to various universality classes. Apart from 
phase transitions, universal quantities associated
with quantum critical regime (QCR) is yet another well-known
example as well \cite{Chu93,Chu931,Chu94,San95,Tro96,Tro97,Tro98,Kim00,Sen15,Tan182}.  
To conclude, the concept of universality 
does play a dominated role in many fields of physics.  

Recently, experimental results of TlCuCl$_3$ \cite{Rue03,Rue08,Mer14} have inspired several
studies of the 3D dimerized spin-1/2 Heisenberg models 
\cite{Kul11,Oit12,Jin12,Kao13,Yan15,Har15,Tan15,Har17,Har171,Tan17,Har172,Tan181}. In particular,
these theoretical investigations have focused on three
universal scaling relations between the N\'eel 
temperature $T_N$ and the staggered magnetization density $M_s$. 
Two of them, namely $T_N/\overline{J}$ versus $M_s$ and
$T_N/T^{\star}$ against $M_s$ will be the main topics
presented in this study. The $\overline{J}$ and
$T^{\star}$ appearing above are the summation of
antiferromagnetic couplings connecting to a spin
and the temperature $T$ at which the uniform susceptibility
$\chi_u$ take its maximum value, respectively.

The universal scaling between $T_N/\overline{J}$ ($T_N/T^{\star}$) 
and $M_s$ is firstly demonstrated in Ref.~\cite{Jin12}. Particularly the
models considered in Ref.~\cite{Jin12} has the property that each spin is
touched by one antiferromagnetic coupling which has larger magnitude
than the rest attaching to the same spin (The antiferromagnetic 
couplings will be called bonds whenever no confusion arises). 
Extending the work of Ref.~\cite{Jin12}, classification schemes 
for both the scaling relations are established in Ref.~\cite{Tan181}. 
Specifically, the scaling relations
between $T_N$ and $M_s$ mentioned above can be categorized
by the number of strong bonds emerging from each spin. 

It is interesting
to notice that in Refs.~\cite{Jin12,Tan181}, all the considered models have two 
kinds of bonds only. 
Moreover, the investigations are carried out on cubic and double-cubic 
lattices which are in a sense both of the same type in geometry. As a result, 
it will be interesting to examine whether the found classification schemes 
are valid for other kinds of lattice geometries, and when additional (spatially) 
anisotropic parameters are introduced into the systems.

Due these intriguing motivations described above, here using the quantum Monte 
Carlo (QMC) simulations, we have studied these two scaling relations of $T_N$ 
and $M_s$
on both the stacked honeycomb and the cubic lattices. Furthermore, a tunable parameter
is taken into account in our investigation so that the studied quantum spin
systems with three types of antiferromagnetic bonds can 
arise.   

While as one expects that the data determined from models with two kinds
of bonds on the stacked honeycomb lattice do fall on
top of the universal curves obtained in Ref.~\cite{Tan181}, mild effects because of
the microscopic details appear. In addition, our results strongly suggest
that a yet to be discovered rule exists since some outcomes from two 
different lattice geometries collapse smoothly to form a curve. Finally,
the most compelling observation implying here is that with the presence of
the new (anisotropic) parameter, the two scaling relations studied
in this investigation can be categorized by the coordinate number of the 
underlying lattices (This will be explained in detail later). This new 
rule can be treated as a very useful supplement to the ones found in Ref.~\cite{Tan181}   

The rest of this paper is organized as follows. After the introduction,
the models as well as the relevant observable are introduced. Following
that we present our results. In particular the numerical evidences for
the new classification rules mentioned above are demonstrated. Finally,
a section concludes our study.

\section{Microscopic models and observables}
\begin{figure}
\vskip-0.5cm
\begin{center}
\vbox{
\hbox{
\includegraphics[width=0.23\textwidth]{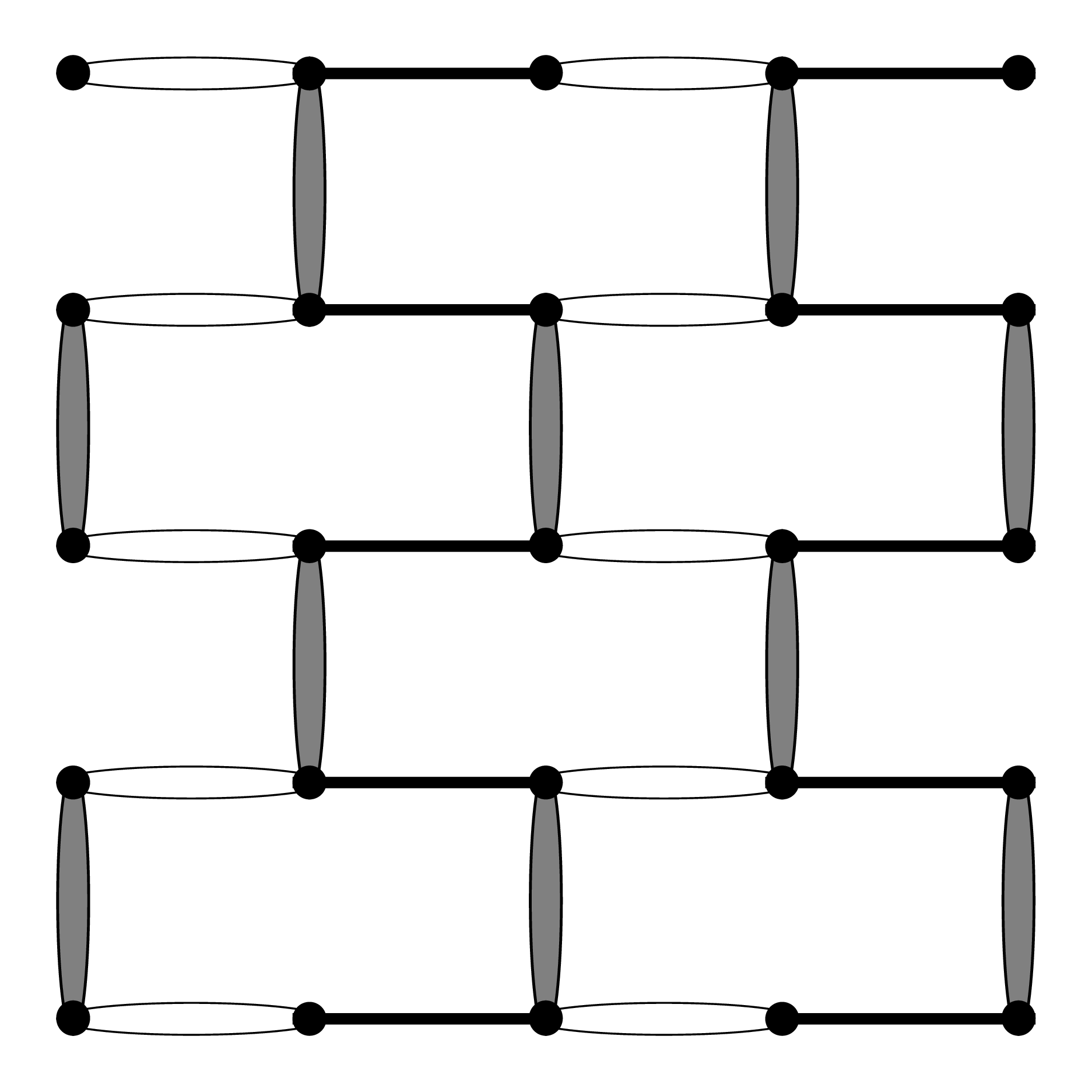}~~~
\includegraphics[width=0.23\textwidth]{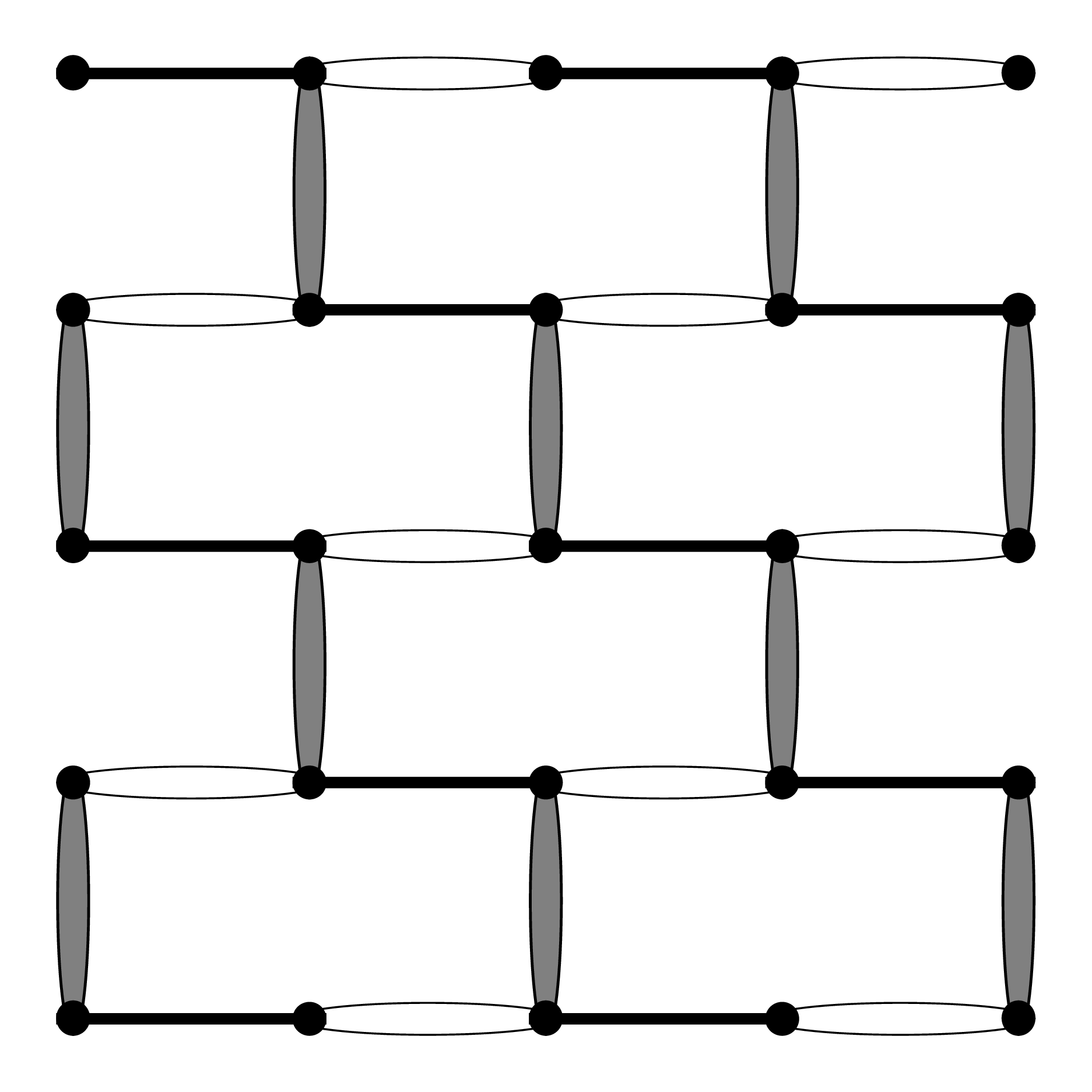}~~~
}\vskip0.75cm
\hbox{
\includegraphics[width=0.23\textwidth]{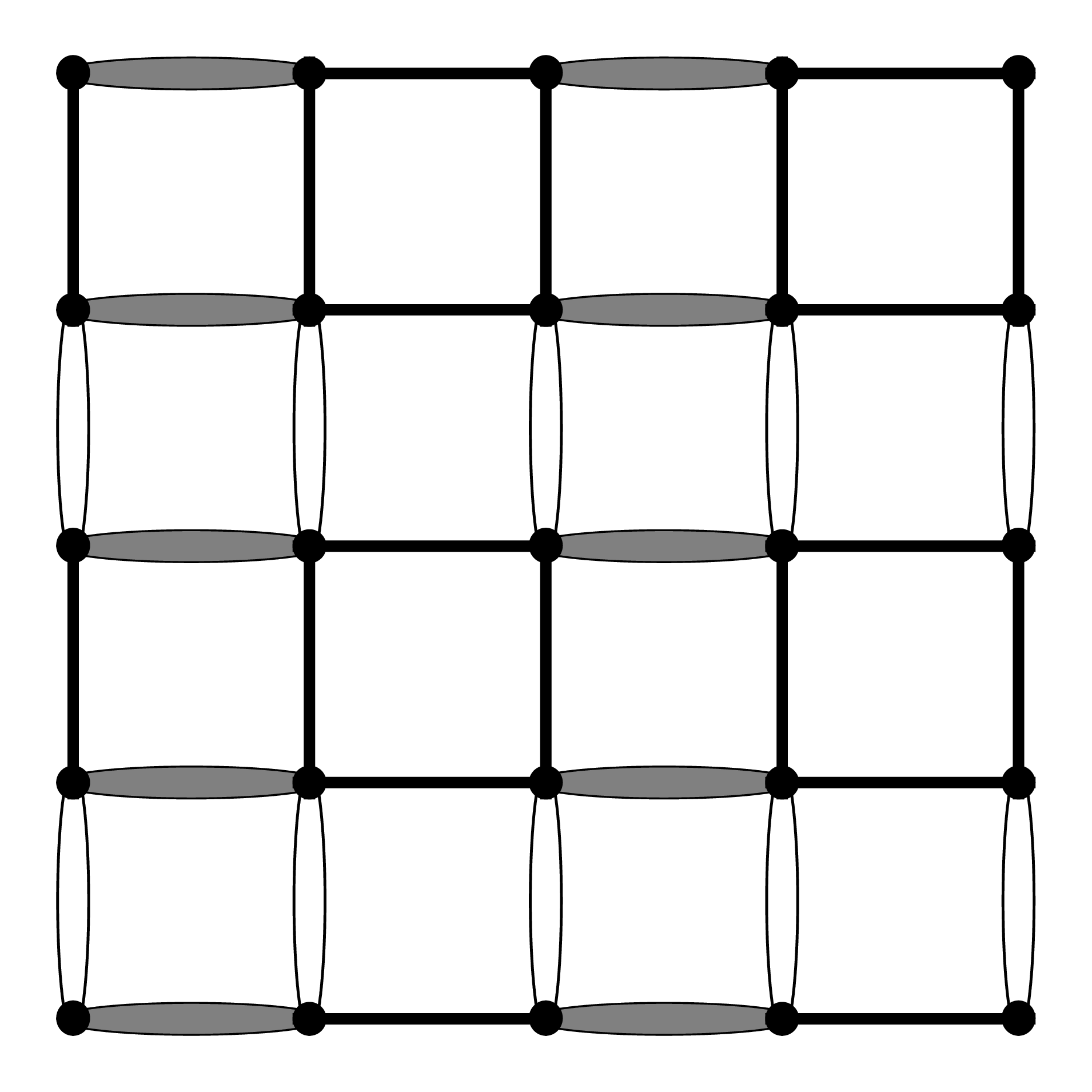}~~~
\includegraphics[width=0.23\textwidth]{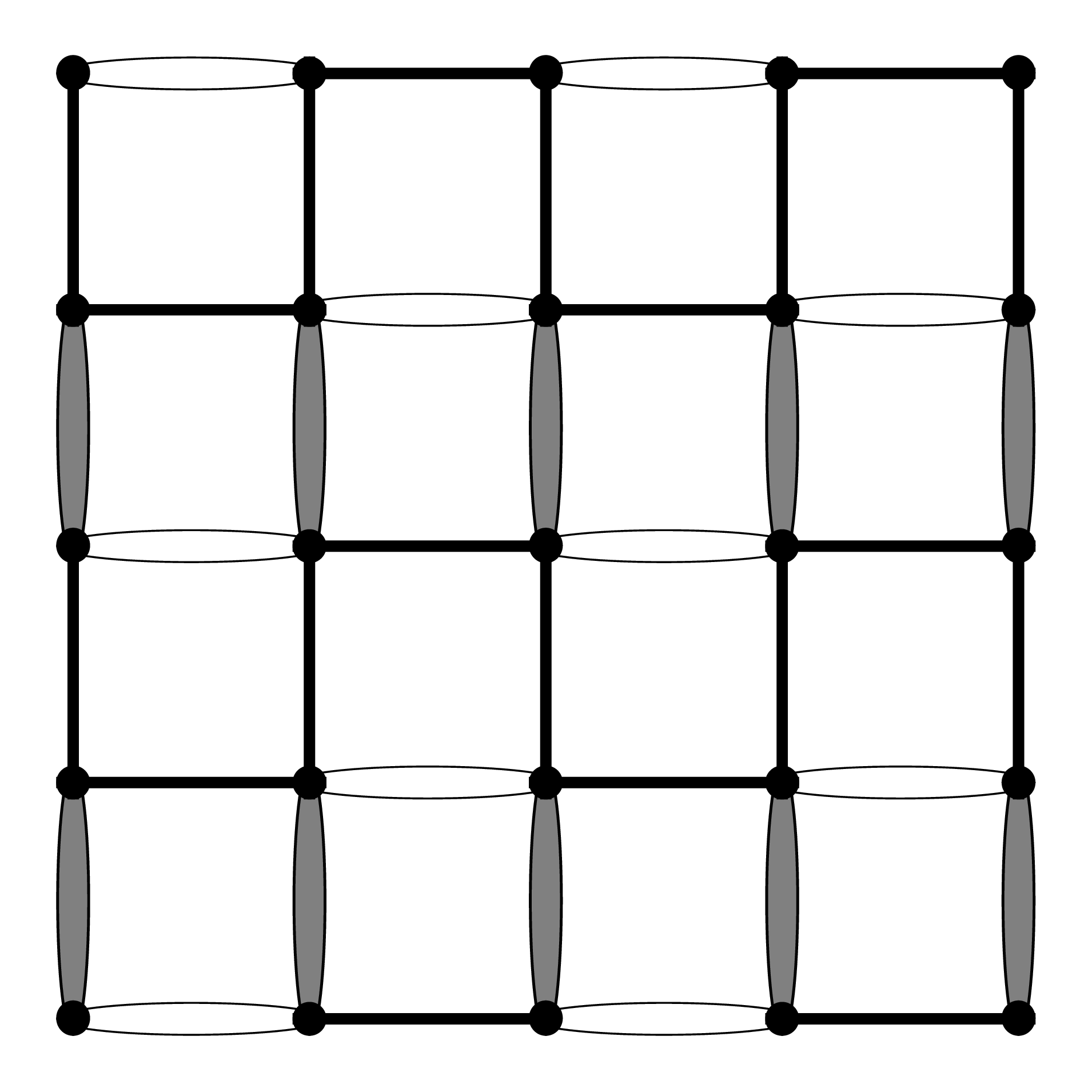}
}}
\end{center}\vskip-0.5cm
\caption{The bond arrangement in the $x$-$y$ plane of the 3D dimerized spin-1/2 
Heisenberg models on the stacked honeycomb and the cubic lattices
investigated here. The filled and empty ovals, 
as well as the thin line represent the bonds with antiferromagnetic couplings $J'$, $\alpha J'$ and $J$, 
respectively. Here $0 < \alpha \le 1$. For all the considered models, 
in the $z$-direction one has only $J$ and $J'$ bonds and they are always set 
up alternately. 
Model on the stacked honeycomb lattice with each spin touching two strong bonds
is obtained by letting $J_{2} = J'$ and $J_{1}=J$ for these bonds in each $x$-$y$ plane. The model of the left top and bottom panels are called the stair and 
meander models here, respectively.}
\label{model_fig1}
\end{figure}

\vskip-0.1cm
The Hamiltonian of the studied 3D spin-1/2 dimerized antiferromagnets on the
stacked honeycomb and the cubic lattices is generally given by
\begin{eqnarray}
\label{hamilton}
H &=& \sum_{\langle ij \rangle}J_{ij}\,\vec S_i \cdot \vec S_{j} 
+ \sum_{\langle i'j' \rangle}\alpha_{i',j'}J'_{i'j'}\,\vec S_{i'} \cdot \vec S_{j'}, 
\end{eqnarray}
where in Eq.~(1) $J_{ij}$ and $J'_{i'j'}$ are the antiferromagnetic 
couplings (bonds) connecting nearest neighbor spins $\langle  ij \rangle$ 
and $\langle  i'j' \rangle$ located at sites of the considered 3D lattices, 
respectively. In addition, for each pair of ${i',j'}$ the associated anisotropic 
factor $\alpha_{i'j'}$ satisfies $ 0 < \alpha_{i'j'} \le 1$. 
Finally, $\vec S_{i} $ is the spin-1/2 operator at site $i$.  
In this study, for any site pairs ${i'j'}$ and ${ij}$, 
we have set $J_{ij} = 1$ and use the convention $J'_{i'j'} > J_{i,j}$. 
Figure~\ref{model_fig1} demonstrates the bond arrangement in the $x$-$y$ plane 
of the dimerized spin-1/2 models studied here. Moreover, for all
the considered systems, in the $z$-direction one has only $J$ and $J'$ bonds and
they are always set up alternately.  
From fig.~\ref{model_fig1} as well as the associated caption, one finds that each spin of the studied 
models is connected to antiferromagnetic couplings of either two or three 
kinds of strength. With the conventions employed here, for each model the 
targeted quantum phase transition is induced by tuning the ratio $J'/J$.

To carry out the proposed investigation, particularly to calculate $T_N$, $M_s$,
as well as $T^{\star}$ of the considered dimerized systems, several observables
including the staggered structure factor $S(\pi,\pi,L)$ on a finite lattice
with linear size $L$ \cite{boxsize}, 
both the spatial and temporal winding numbers squared 
($\langle W_i^2 \rangle$ for $i \in \{1,2,3\}$ and
$\langle W_t^2 \rangle$), spin stiffness $\rho_s$,
first Binder ratio $Q_1$, and second Binder ratio $Q_2$  
are measured. The definitions of these physical quantities as well
as how they can be recorded in the associated
Monte Carlo simulations are well known and are available in 
numerous relevant publications, see Ref.~\cite{San97} for
a detailed introduction.

The staggered structure factor $S(\pi,\pi,\pi,L)$, which is relevant for 
the determination of $M_s$, is defined by 
\begin{equation}
S(\pi,\pi,\pi,L) = 3 \langle ( m_s^z )^2\rangle,
\end{equation}
where $m_s^z = \frac{1}{L_1L_2L_3}\sum_{i}(-1)^{i_1+i_2+i_3}S^z_i$.
Here $S^z_i$ is the third component of the spin-1/2
operator $\vec S_i$ at site $i$. Furthermore,
the spin stiffness $\rho_s$ is calculated through
\begin{equation}
\rho_s = \frac{1}{3}\sum_{i=1,2,3} \rho_{si} = \frac{1}{3\beta}\sum_{i=1,2,3}\frac{\langle W_i^2 \rangle}{L_i},
\end{equation}
where $\beta$ is the inverse temperature, and $W_i$ with $i\in\{1,2,3\}$
are the spatial winding numbers. Besides these observables,
the temporal winding number squared $\langle W_t^2\rangle$, which is expressed as
\begin{eqnarray}
\langle W_t^2\rangle = \left\langle \left(\sum_i S^z_i\right)^2 \right\rangle ,
\end{eqnarray}
is calculated in our study as well.
Finally the observables $Q_1$ and $Q_2$ are defined by
\begin{equation}
Q_1 = \frac{\langle |m_s^z| \rangle^2 }{\langle (m_s^z)^2\rangle} 
\end{equation}
and
\begin{equation}
Q_2 = \frac{\langle (m_s^z)^2 \rangle^2}{\langle (m_s^z)^4\rangle},
\end{equation} 
respectively.

\begin{figure}
\begin{center}
\includegraphics[width=0.4\textwidth]{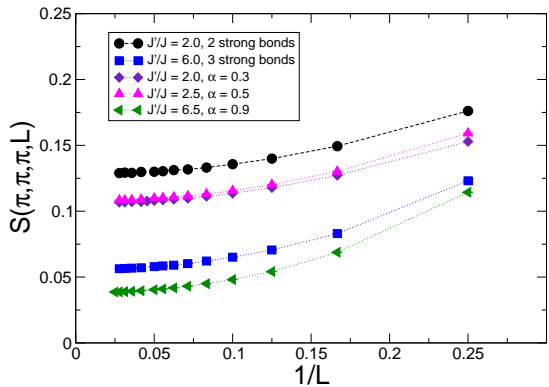}
\end{center}\vskip-0.2cm
\caption{The $1/L$ dependence of the staggered structure factors 
$S(\pi,\pi,\pi,L)$ for 
some of the considered 3D quantum spin models on the stacked honeycomb lattice.
The anisotropic factor $\alpha$ and $J'/J$ associated with each of the presented data 
sets are specified in the legend. The dashed lines are added to guide the eye.}
\label{ms_fig1}
\end{figure}

\vskip0.5cm

\section{The numerical results}
\vskip-0.1cm
To investigate the $\alpha$ dependence of the scaling relations between $T_N$ and $M_s$, particularly to understand how these universal 
curves shown in 
Refs.~\cite{Tan181} get modified, we have carried out a large-scale QMC 
simulation using the stochastic series expansion (SSE) algorithm with very 
efficient operator-loop update \cite{San99}. To begin with, in the following 
we will firstly present our determination of $M_s$.


\subsection{The determination of $M_s$}

For each considered value of $J'/J$, the associated $M_s$ can be derived from 
$S(\pi,\pi,\pi,L)$ obtained at zero temperature by 
$\sqrt{S(\pi,\pi,\pi,L \rightarrow \infty)}$. 
Here the zero temperature results of $S(\pi,\pi,\pi,L)$ are reached through
simulations using $\beta = 2L$. We would like to point out that for small to 
intermediate lattices, larger $\beta$ than $\beta = 2L$ are employed. 
For several studied models and some selected $J'/J$, we have additionally performed a few simulations 
using $\beta > 2L$ (including those done with 
$\beta = 4L$). The results obtained from these trial calculations agree very 
well with those explicitly presented in this investigation. 
Therefore, the determined $M_s$ shown here should be the ones corresponding to 
the ground states. 

\begin{figure}
\begin{center}
\includegraphics[width=0.4\textwidth]{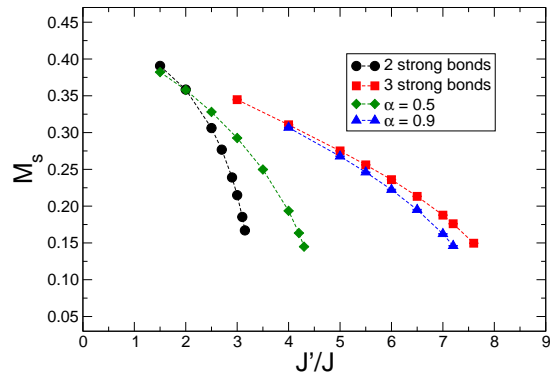}
\end{center}
\vskip-0.3cm
\caption{$M_s$ as functions of the considered $J'/J$ for several 
of the studied 3D quantum spin models on the stacked honeycomb lattice. 
The anisotropic factor $\alpha$ associated with each of the presented data 
sets are specified in the legend. The dashed lines are added to guide the eye.}
\label{ms_fig2}
\end{figure}
  
For several studied models, the $1/L$-dependence of their ground 
states $S(\pi,\pi,\pi,L)$ for some considered $J'/J$ are depicted 
in figs.~\ref{ms_fig1}. Following Refs.~\cite{Car96}
the numerical values of $M_s$ are obtained by 
performing extrapolations in $1/L$ using the following three ansatzes
\begin{eqnarray}
\label{poly2}&&a_0 + a_2/L^2, \\
\label{poly3}&&b_0 + b_2/L^2 + b_3/L^3, \\
\label{poly4}&&c_0 + c_2/L^2 + c_3/L^3 + c_4/L^4,
\end{eqnarray}
In particular, the corresponding results of $M_s$ are determined by taking the 
square roots of $a_0, b_0, c_0$ calculated from the fits. In some cases,
formulas up to fifth order in $1/L$ are used for the fits. The calculated
numerical values of $M_s$ for the studied models are shown in fig.~\ref{ms_fig2}.
The data presented in that figure are obtained by averaging over all
the good fits (Which are defined as those with a $\chi^2/{\text{DOF}} < 2.0$).
Furthermore, for every studied model and for each considered parameter $J'/J$, 
the corresponding uncertainty shown in the figure is based on 
the associated errors from all the (good) fits related to it.

\subsection{The determination of $T_N$}

\begin{figure}[h]
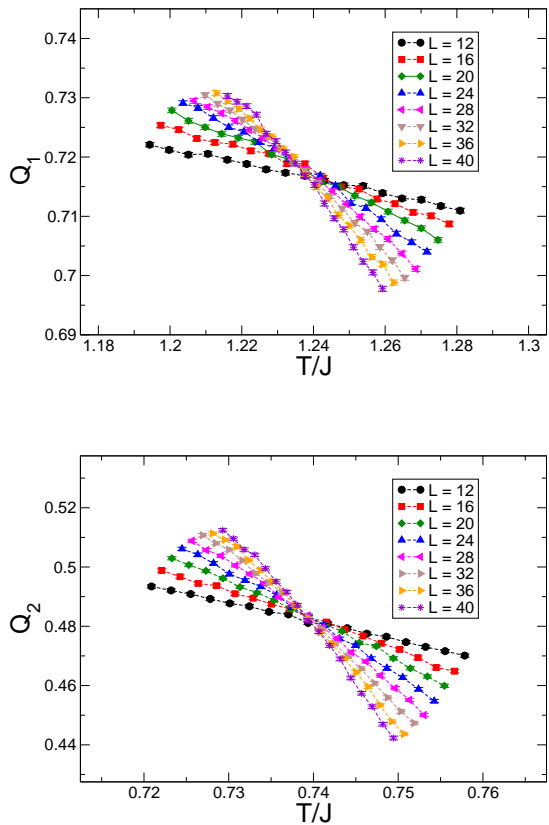

\vskip0.5cm
\begin{center}
\vbox{
\includegraphics[width=0.4\textwidth]{3bonds_Q1_J6.0.eps}\vskip1cm
\includegraphics[width=0.4\textwidth]{2bonds_Q2_J2.7.eps}
}
\end{center}
\caption{$Q_1$ (top panel, three strong bonds, $J'/J = 6.0$) and $Q_2$ 
(bottom panel, two strong bonds, $J'/J = 2.7$) as functions of $T/J$ for 
various $L$. The dashed lines are added to guide the eye.}
\label{TN_fig1}
\end{figure}

The N\'eel temperatures $T_N$ for various $J'/J$ of the studied models are 
calculated by applying the expected finite-size scaling to the relevant
observables. Specifically, $T_N$ are determined through bootstrap-type fits 
using constrained standard finite-size scaling ansatz of the form 
\begin{eqnarray}
(1+b_0L^{-\omega})(b_1 + b_2tL^{1/\nu} +b_3(tL^{1/\nu})^2+...).
\end{eqnarray}
Here $b_i$ for $i=0,1,2,...$ are some constants and $t = \frac{T-T_N}{T_N}$.
Moreover, this ansatz with up to second, third, fourth order 
and (or) fifth order in $tL^{1/\nu}$ are carried out to fit the data
of $Q_1$, $Q_2$ and $\rho_s L$. The $Q_1$ ($Q_2)$ data of one of the 
investigated 
models are shown in the top (bottom) panel of fig.~\ref{TN_fig1}. 

\begin{figure}[h]
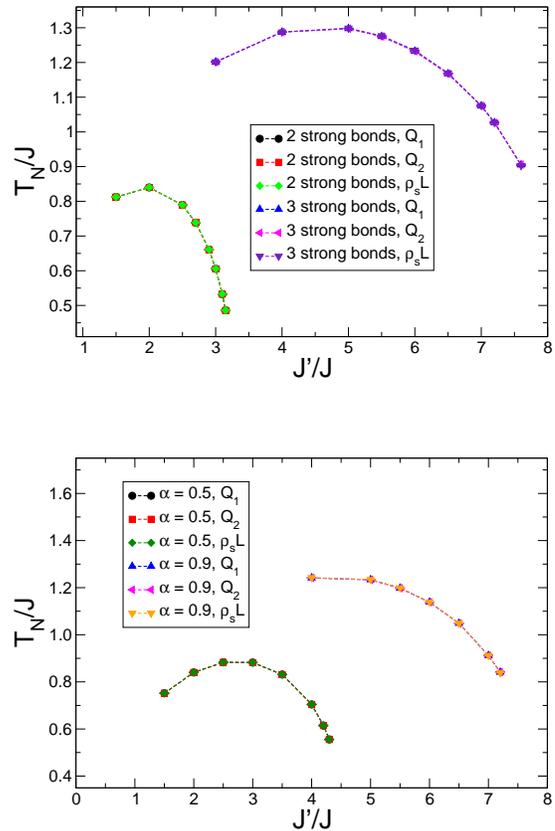

\vskip0.5cm
\begin{center}
\vbox{
\includegraphics[width=0.4\textwidth]{TN_Q1_Q2_rhosL_2_3_bonds.eps}\vskip1cm
\includegraphics[width=0.4\textwidth]{TN_Q1_Q2_rhosL_aniso.eps}\vskip1cm
}
\end{center}
\caption{The $J'/J$ dependence of $T_N$ obtained from $Q_1$, $Q_2$,
and $\rho_sL$ for some considered 3D spin models studied here. 
The $\alpha$ corresponding to the data presented in the 
figure are shown explicitly in the legend.
The dashed lines are added to guide the eye.}
\label{TN_fig2}
\end{figure}

For the considered models, the detailed steps of estimating the
corresponding $T_N$ including their associated uncertainties are the 
same as those demonstrated in Ref.~\cite{Tan181}. With the procedures 
describing in Ref.~\cite{Tan181}, the $T_N$ obtained from the three used 
observables for several of the studied models are shown in figs.~\ref{TN_fig2}.
It should be pointed out that for some cases, while the $T_N$ determined from 
considering the observable $\rho_s L$ are slightly different from those 
related to $Q_1$ and $Q_2$, the variations are merely at few per mille level. 
Therefore, one expects that such small discrepancies have no influence on the 
conclusions obtained here.    

\vskip0.25cm

\begin{figure}[h]
\vskip1.0cm
\begin{center}
\includegraphics[width=0.4\textwidth]{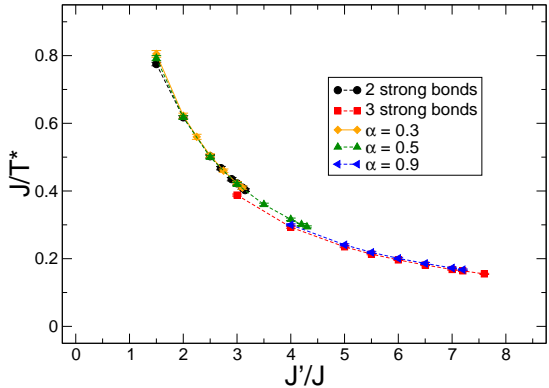}\vskip1cm
\end{center}
\caption{The inverse of $T^{\star}$ as functions of $J'/J$ for some 3D spin 
models studied here. The $\alpha$ corresponding to the data presented in the 
figure are shown explicitly in the legend.
The dashed lines are added to guide the eye.}
\label{Tstar_fig1}
\end{figure}  

\subsection{The determination of $T^{\star}$}

For all the investigated models, the temperatures at which $\chi_u$ reach their
maximum value (These temperatures are denoted by $T^{\star}$) are determined on 
lattices with $L=16$. The estimations of the inverse of $T^{\star}$ as 
functions of $J'/J$ for several considered systems are shown in 
fig.~\ref{Tstar_fig1}.

For some models, simulations with $L=32$ are conducted in order to 
understand the effects of finite-size on the determination of $T^{\star}$.
For these additional calculations we find that the results obtained on 
$L=16$ lattices are already the bulk ones. Based on these studies on 
$L = 32$ lattices as well as those presented in Refs. \cite{Tan181}, it is 
anticipated the conclusions obtained in the following (sub)section by employing these 
estimated $T^{\star}$ (on $L=16$ lattices) should be reliable.

\subsection{The scaling relations between $T_N/\overline{J}$, $T_N/T^{\star}$, 
and $M_s$}

\begin{figure}
\vskip0.5cm
\begin{center}
\includegraphics[width=0.4\textwidth]{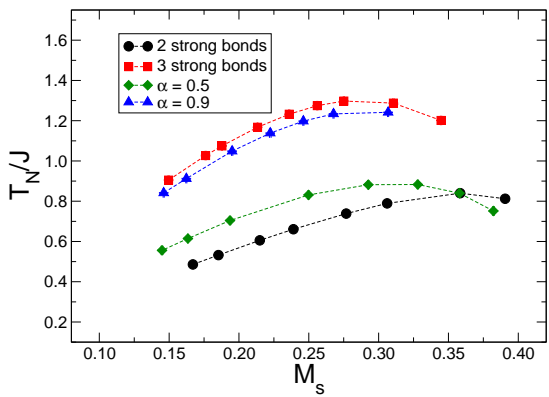}
\end{center}
\caption{$T_N/J$ as functions of $M_s$ for several considered 3D dimerized 
models on the stacked honeycomb lattice.
Each $T_N$ shown in the figure is obtained from $Q_1$.
The dashed lines are added to guide the eyes.}
\label{TN_barJ_Tstar_fig1}
\end{figure}

\begin{figure}
\vskip1.0cm
\begin{center}
\includegraphics[width=0.4\textwidth]{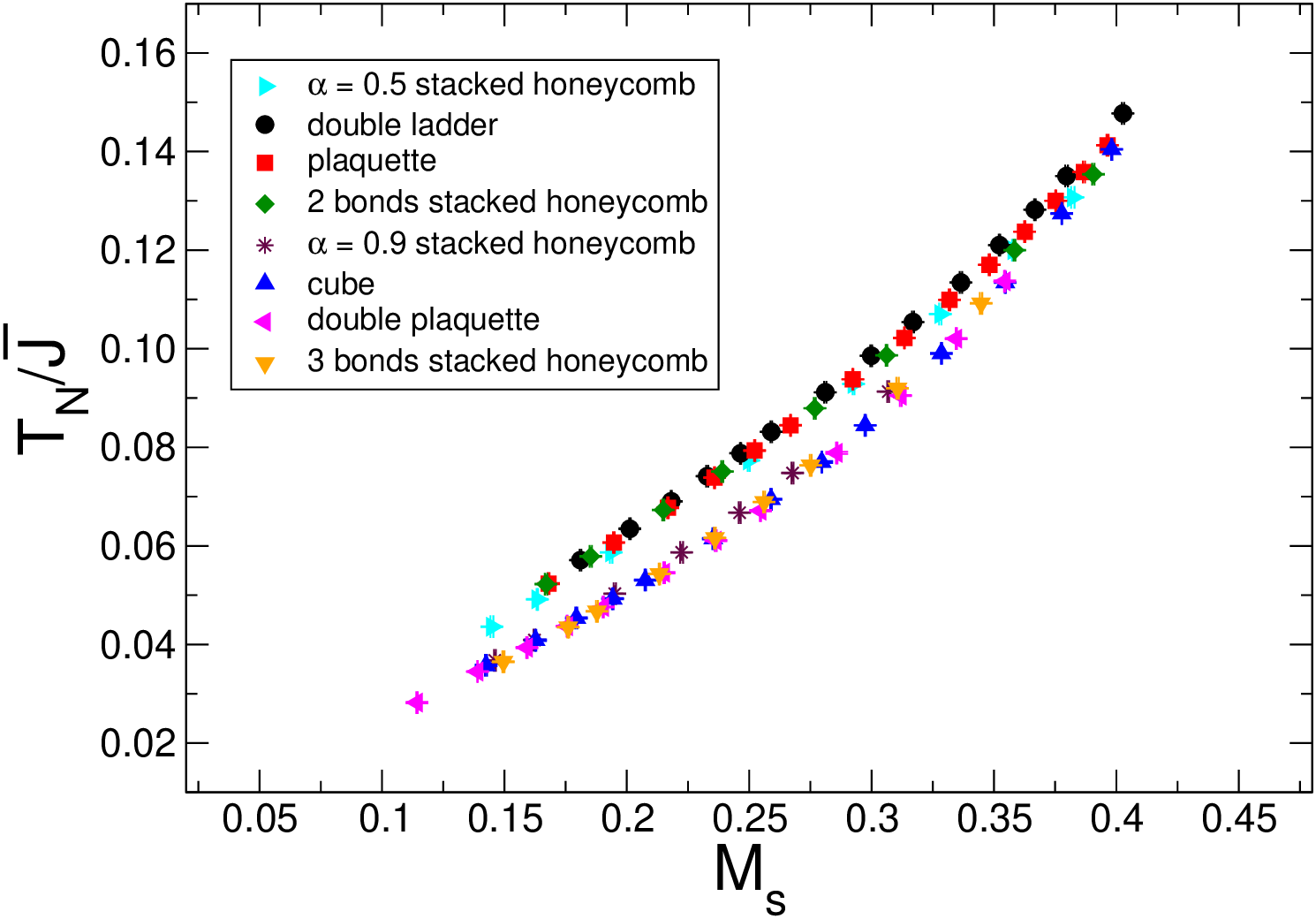}
\end{center}
\caption{$T_N/\overline{J}$ as functions of $M_s$ 
for most of the considered models in this study. Each used $T_N$ in the figure
is obtained from $Q_1$. For comparison purpose, 
some data presented in Ref.~\cite{Tan181} are shown in the figure as well.}
\label{TN_barJ_Tstar_fig2}
\end{figure}

\begin{figure}
\vskip0.5cm
\begin{center}
\includegraphics[width=0.4\textwidth]{Universal_TN_Tstar_ms.eps}
\end{center}
\caption{$T_N/T^{\star}$ as functions of $M_s$ for most of the considered models in this study.
Each used $T_N$ in the figure is obtained from $Q_1$. For comparison purpose, 
some data presented in Ref.~\cite{Tan181} are shown in the figure as well.}
\label{TN_barJ_Tstar_fig3}
\end{figure}

\begin{figure}
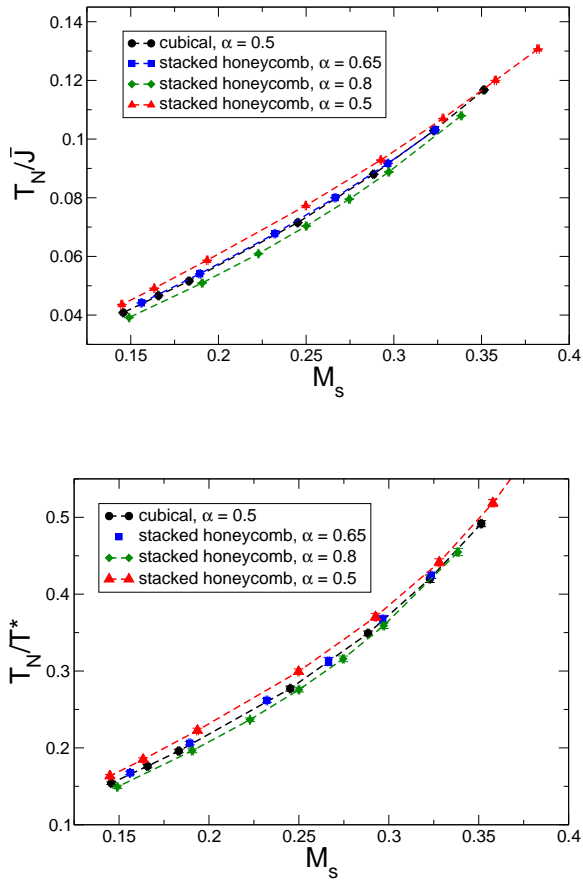

\vskip0.5cm
\begin{center}
\vbox{
\includegraphics[width=0.425\textwidth]{barJ_cubical_0.5_honeycomb_0.65.eps}\vskip1cm
\includegraphics[width=0.425\textwidth]{Tstar_cubical_0.5_honeycomb_0.65.eps}
}
\end{center}
\caption{(Top panel) Universal curves associated with $T_N/\overline{J}$ and $M_s$ for models on the cubic ($\alpha = 0.5$) 
and the stacked honeycomb ($\alpha = 0.5,0.65,0.8$) lattices.
(Bottom panel) Universal curves associated with $T_N/T^*$ and $M_s$ for models on the cubic ($\alpha = 0.5$) 
and the stacked honeycomb ($\alpha = 0.5,0.65,0.8$) lattices. 
The dashed lines are added to guide the eye. For better visualization, in the bottom panel 
the dashed line associated with $\alpha = 0.65$ (stacked honeycomb) is not shown explicitly.}
\label{TN_barJ_Tstar_fig4}
\end{figure}

Using the $M_s$ and $T_N$ determined in the previous 
subsections, 
we find that no clear connections between the curves of $T_N/J$ against $M_s$
among these studied spin-1/2 models on the stacked honeycomb lattice, see
fig.~\ref{TN_barJ_Tstar_fig1} for the outcomes of several considered systems. 
Interestingly, if $T_N/\overline{J}$ are plotted as functions
of $M_s$, universal scaling relations emerges, as can been seen in
fig.~\ref{TN_barJ_Tstar_fig2}. Remarkably, while for systems of 
$\alpha = 0.3$ (not shown in fig.~\ref{TN_barJ_Tstar_fig2}) and $0.5$, as well
as model with two strong bonds attaching to each of its spin, 
the resulting data of $T_N/\overline{J}$ as functions of $M_s$
do form a universal curve, this universal curve falls
on top of the one corresponding to the models investigated in 
Ref.~\cite{Tan181} which have two strong bonds connected to each of
their spin. Similar situation occurs for models with 
$\alpha = 0.9$ and that having three strong bonds emerging from each spin,
namely all of their associated data form a single curve. 
In particular, this universal curve matches the one related to
the cubic and double plaquette models investigated in Ref.~\cite{Tan181}.
Finally, we would like to emphasize the fact that the data
obtained for other values of $\alpha$ here indicate  
as the magnitude of $\alpha$ increases from 0.5 to 0.9, the resulting curves begin
to deviate from the curve associated with two strong bonds and eventually
collapse with the curve corresponding to three strong bonds.  

By considering $T_N/T^{\star}$ as functions of $M_s$ for all the models 
studied here as well as those investigated in Ref.~\cite{Tan181}, the same 
scenario as that of $T_N/\overline{J}$ versus $M_s$ also appears, see 
fig.~\ref{TN_barJ_Tstar_fig3}.

It is remarkably that the classification schemes which are firstly pointed out 
in Ref.~\cite{Tan181} and are valid for cubic 
type lattices are now extended to include 3D quantum spin models on the stacked honeycomb lattice.
While this is the case, effects due 
to microscopic details, specially those of the quantum fluctuations, do 
have minor impact on the
categorization of the universal curves. Indeed, as can be seen from
figs.~\ref{TN_barJ_Tstar_fig2} and \ref{TN_barJ_Tstar_fig3}, when the magnitude of $M_s$ increases, 
the curve related to the models of three strong bonds and $\alpha = 0.9$ 
studied here begins to move toward the curve associated with two strong bonds 
at a value $M_s$ slightly smaller than that of the curve resulting
from the models considered in Ref.~\cite{Tan181}. Since the stacked honeycomb 
lattice has five coordinate number which is fewer than those of the cubic and 
the double cubic lattices, the resulting quantum fluctuation is more profound and have 
greater influence on properties of the systems on the stacked honeycomb lattice. 
Nevertheless,  
it is beyond doubt that the classification schemes claimed in 
Ref.~\cite{Tan181} are valid not only on cubic-type lattices, but also for 
models on the stacked honeycomb 
lattice. In particular, the universal curves associated with the stacked 
honeycomb lattice match those related to the cubic-type lattices.

We would like to point out that while intuitively one expects 
the curve related to a particular value of $\alpha$ will start to
move away from the one of two strong bonds, it in intriguing that
this particular $\alpha$ is larger than (equal to) $0.5$ for the models 
on the stacked honeycomb lattice studied here.   

Apart from quantum spin models on the stacked honeycomb lattice,
we have additionally simulated the cubic model studied in Ref.~\cite{Tan181}.
In particular, an anisotropic bond similar to the $\alpha$-bond considered
here is introduced in our investigation so that models with three types of bonds
can be obtained, see the left bottom panel of fig.~\ref{model_fig1}. 
This generalized
model will be called anisotropic cubic model. Remarkably,
the $T_N/\overline{J}$ and $T_N/T^{*}$ versus $M_s$ data for $\alpha = 0.5$ 
on the anisotropic cubic model fall on the same curve as that of the 
model on the stacked honeycomb lattice with $\alpha = 0.65$, see 
fig.~\ref{TN_barJ_Tstar_fig4}. 
The associated data for $\alpha = 0.8$ and $\alpha = 0.5$ of the models on 
the stacked honeycomb lattice are also shown in fig.~\ref{TN_barJ_Tstar_fig4}. 
With these two additional sets of data, one can sees clearly 
the good data collapse quality from both the systems on the 
anisotropic cubic lattice with $\alpha = 0.5$ and on the
stacked honeycomb lattice with $\alpha = 0.65$. The outcomes demonstrated in 
both top and bottom
panels of fig.~\ref{TN_barJ_Tstar_fig4} suggest convincingly that there is 
yet a to be understood categorization rule for the anisotropic models with 
$0 < \alpha < 1$.     

\begin{figure}
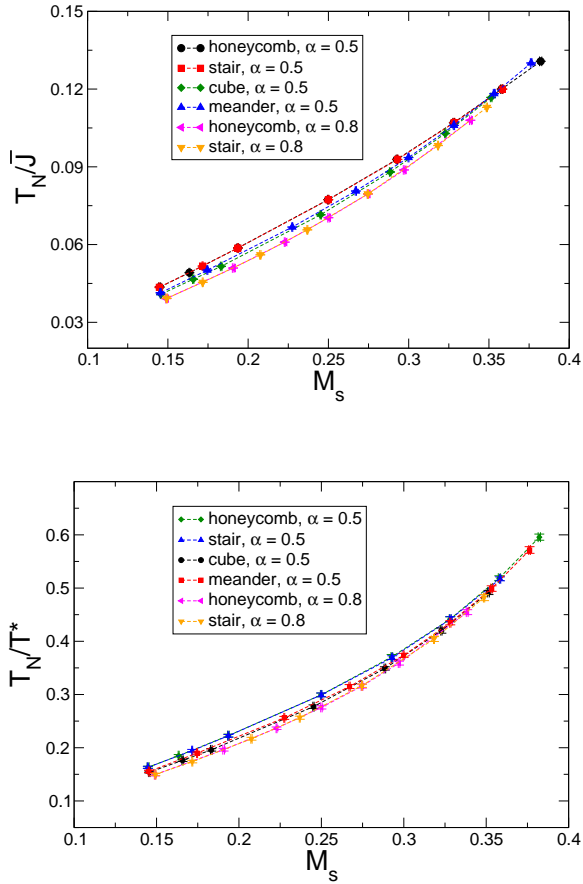

\vskip0.5cm
\begin{center}
\vbox{
\includegraphics[width=0.425\textwidth]{universal_anisotropic_barJ_J0.5.eps}\vskip1cm
\includegraphics[width=0.425\textwidth]{universal_anisotropic_Tstar_J0.5.eps}
}
\end{center}
\caption{(Top panel) Universal curves associated with $T_N/\overline{J}$ and 
$M_s$ for models on both the cubic ($\alpha = 0.5$)
and the stacked honeycomb ($\alpha = 0.5,0.8$) lattices. (Bottom panel) Universal 
curves associated with $T_N/T^*$ and $M_s$ for 
models on both the cubic ($\alpha = 0.5$) and the stacked honeycomb 
($\alpha = 0.5,0.8$) lattices. The dashed lines are added to guide the eyes.}
\label{TN_barJ_Tstar_fig5}
\end{figure}

Besides the results presented above, another compelling outcome from our 
investigation is that for both lattice geometries,
data collapse of $T_N/\overline{J}$ ($T_N/T^{\star}$) versus $M_s$ with $\alpha = 0.5$ within each category of lattice geometries  
(and $\alpha = 0.8$ on the stacked honeycomb lattice)
lead to a smooth curve, see both panels of fig.~\ref{TN_barJ_Tstar_fig5}. 
Specifically, the curves of related data of both models on the top (bottom) panel of 
fig.~1
with $\alpha = 0.5$ ($\alpha = 0.5$), which are different models on the stacked honeycomb (cubic) lattice, 
fall on top of each other.
The situation also occurs for models of $\alpha = 0.8$ associated
with the stacked honeycomb lattice, but its universal curve differs from the one
of $\alpha = 0.5$. Based on these outcomes, it is highly probable that such 
a scenario occurs for other values of $\alpha$. This result strongly suggests 
that for each value of spatial anisotropy, the universal characteristics 
between $T_N$ and $M_s$, which were found in Refs. \cite{Jin12,Tan181} can be 
classified by the coordinate number of the underlying lattice geometries. 
This observation for 3D anisotropic quantum spin systems is new, and was not 
established before in the literature. We would like to emphasize the fact that 
in fig.~\ref{TN_barJ_Tstar_fig5} the quality of data collapse for the two 
different models on the stacked honeycomb lattice is much better than those of the 
systems on the cubic lattice. It might be interesting to understand this 
result from a theoretical point of view.   

Finally, it should be pointed out that although $\overline{J}$ and 
$T^{\star}$ are two completely different quantities, it is remarkable that 
based on the results presented in Refs.~\cite{Jin12,Tan181} and here, 
the categorization schemes for $T_N/\overline{J}$ versus $M_s$ and 
$T_N/T^{\star}$ versus $M_s$ are totally identical to each other. This 
implies there may be an even more fundamental classification principle than 
those already explored.

\section{Discussions and Conclusions}
\vskip-0.1cm

Using the first principles quantum Monte Carlo simulations, we have 
investigated in detail the universal scaling relations between $T_N$ and 
$M_s$,  namely $T_N/\overline{J}$ versus $M_s$ and $T_N/T^{\star}$ versus 
$M_s$ for 3D quantum antiferromagnets on both the stacked honeycomb and
the cubic lattices.   

By studying the 3D spin-1/2 dimerized Heisenberg models with two types of
antiferromagnetic coupling strength, in Ref.~\cite{Tan181} it was established 
that these universal relations can be classification by the number of 
$J'$-bonds touching each spin of the considered models. Here we extend these 
categorization schemes by investigating systems with three kinds of bonds and
on lattices of different geometries. 

According to the outcomes obtained here, while the classification rules for
anisotropic cases are more complicated than the ones established in 
Ref.~\cite{Tan181},
without doubt a generalized categorization principle does exist for these
models with $0 < \alpha < 1$. Particularly, we conjecture that with the 
presence of three types of bonds and for a given $\alpha$, the categorization 
rule is in accordance with the coordinate number of the underlying lattice 
geometry. More surprisingly, although $T^{\star}$ and $\overline{J}$ are two 
completely different physical quantities, the classification schemes for 
these two relations are identical. 

To understand the theories relevant to the results obtained here, particularly 
to uncover the corresponding mechanism behind the identical classification 
schemes for two different universal relations observed in this study, 
will definitely be interesting and compelling to pursue in the future.


\section{Acknowledgments}
\vskip-0.5cm
This study is partially supported by MOST of Taiwan.


\begin{thebibliography}{99}
\bibitem{boxsize}
Most of the simulations are done with $L_1 = L_2 = L_3$. Here we use $L$
to stand for $L_1$ unless specified.
\end{thebibliography}
\end{document}